\documentclass[]{interact}
\usepackage{ragged2e}
\usepackage{apacite}

\usepackage{enumerate}
\usepackage[shortlabels]{enumitem}

\usepackage{epstopdf}
\usepackage[caption=false]{subfig}

\usepackage{comment}

\makeatletter
\def\xfnm[#1]{%
  \def\xfnmarg{#1}\def\xfnmrelax{ {\relax}.}%
  \ifx\xfnmarg\xfnmrelax
    \expandafter\@firstoftwo
  \else
    \expandafter\@secondoftwo
  \fi
  {,\@gobble}%
  {\unskip,\space#1}%
}
\makeatother

\usepackage{xcolor}

\usepackage[sort]{natbib}
\bibpunct[, ]{(}{)}{;}{a}{,}{,}
\renewcommand\bibfont{\fontsize{10}{12}\selectfont}

\theoremstyle{plain}

\theoremstyle{definition}

\theoremstyle{remark}

\begin{document}


\title{Towards augmented reality for corporate training}

\author{
\name{Blind for review.}
}


\author{
\name{B.~R. Martins\textsuperscript{a,b}, 
J.~A. Jorge\textsuperscript{c} and E.~R. Zorzal\textsuperscript{a,c}\thanks{CONTACT E.~R. Zorzal. Email: ezorzal@unifesp.br}}
\affil{\textsuperscript{a}Instituto de Ciência e Tecnologia, Universidade Federal de São Paulo; \textsuperscript{b}Embraer S.A., Brazil; \textsuperscript{c}Instituto Superior Técnico, Universidade de Lisboa, INESC-ID Lisboa, Portugal}
}


\maketitle

\begin{abstract}
Corporate training relates to employees acquiring essential skills to operate equipment or effectively performing required tasks both competently and safely. Unlike formal education, training can be incorporated in the task workflow and performed during working hours. Increasingly, organizations adopt different technologies develop both individual skills and improve their organization. Studies indicate that Augmented Reality (AR) is quickly becoming an effective technology for training programs. This systematic literature review (SLR) aims to screen works published on AR for corporate training. We describe AR training applications, discuss current challenges, literature gaps, opportunities, and tendencies of corporate AR solutions. We structured a protocol to define keywords, semantics of research, and databases used as sources of this SLR. From a primary analysis, we considered 1952 articles in the review for qualitative synthesis. We selected 60 among the selected articles for this study. The survey shows a large number of 41.7\% of applications focused on automotive and medical training. Additionally, 20\% of selected publications use a camera-display with a tablet device, while 40\% refer to head-mounted-displays, and many surveyed approaches (45\%) adopt marker-based tracking. Results indicate that publications on AR for corporate training increased significantly in recent years. AR has been used in many areas, exhibiting high quality and provides viable approaches to On-The-Job training. Finally, we discuss future research issues related to increase relevance regarding AR for corporate training.
\end{abstract}



\section{Introduction}
Manufacturing companies face serious challenges related to ever-changing demands by customers and suppliers alike. New technological changes and suggested interventions aim to exploit the economic potential resulting from rapidly advancing information and communication technology in the industry. Training the workforce is not a one-time issue to be addressed. Instead, it is an ongoing effort that must be nurtured and integrated into the corporate culture.


Indeed, it must continuously be on the managers' table. 
Certainly, on-the-job training has the potential to both improve worker productivity and save resources. Today we witness what could be the dawn of the fourth industrial revolution coming on the heels of information technologies (IT). This revolution is characterised by Internet technologies becoming pervasive. Indeed, IT is becoming easier to use and permeate our everyday lives via intelligent components, robots, the internet of things (IoT), and interactive technologies. 

These developments lie at the core of emerging smart factories where physical and digital systems become integrated to enable both mass customisation and faster product development \citep{Demartini2017}. Also, industries are investing in new technologies to improve complex processes \citep{blanco2018practical}. According to Cardoso et al. \cite{zorzal2020}, Augmented Reality (AR) is one of the leading technologies in this context.
This AR lead happens because it can be applied in different industrial environments to improve process flexibility, providing information to manufacturing, improving product inspection, and providing more efficient logistics while supporting maintenance process monitoring.


AR is currently used in many different niche applications and diverse goals such as emulating critical situations with low human risks, creating promising opportunities to train medical professionals in a safe environment, providing general academic resources, serving as maintenance facilitator devices, in the manufacturing industries such as automotive and aeronautical as well.

AR is known as a technology that allows computer-generated virtual imagery to overlay physical objects in real-time accurately \citep{zhou:hal-01530565}. In AR most visual sensations and sensory stimuli come from the real world, and the virtual elements contribute less, the virtual object is set as if it were part of the real world. In Virtual Reality (VR), most sensory information is computer-generated where the virtual environment is presented as if the user were part of it. 

AR applications are becoming even more affordable, thanks to more powerful hardware, including processors, head-mounted-displays (HMD) and smaller form-factors such as smartphones. Thanks to these progresses and even more sophisticated user interfaces AR is mounting to new levels of usability \citep{wanderley2006survey}.

AR, as an interaction paradigm, is predicted to be one of the enabling technologies that will power the transformation supported by the Industry 4.0 initiative \citep{Davies2015}, which is expected to revolutionise the current production systems. Indeed, AR is welcome in manufacturing \citep{Damiani2018} as it can help humans to:

\begin{enumerate}
 \item Speed up reconfiguration of production lines;
 \item Support shop-floor operations; 
 \item Implement virtual training for assembling parts;
 \item Manage the warehouse efficiently;
 \item Support advanced diagnostics integrated into modules with the working environment \citep{Damiani:2015:ARS:2888619.2889048}.
\end{enumerate}

This SLR focuses on the topic: Implementing virtual training for assembling parts. It is not circumscribed to industry settings but also delves on the medical, service, military and many other corporate training applications. Our goal is to elaborate a comprehensive SLR on AR as a training paradigm focused providing just-in-time~\cite{MAR2005} instructions to people so that they can perform activities more efficiently with minimum supervision from senior staff and with online assessment so as to become a reliable foundation for On-the-Job Training (OJT).

In the context of this survey we look for training suites with the potential to be used for OJT. This requires both the ability to update pedagogical content on short notice and, more importantly, to integrate the module in the actual task workflow.
OJT makes it possible to offer spontaneous explanations or demonstrations related to a person's job responsibilities and performance requirements.  
Proper OJT enables people to hone their skills either by trial-and-error learning or by observing and imitating the behaviours of others \citep{jacobs1999status}. Although not all the studies selected in this survey target OJT, all show potential to be used in that context.

As both service providers and manufacturing systems switch from mass production to mass customisation \citep{Burger2017}, this leads to more client-specific, tailor-made products and services delivered to an even diverse and larger customer base.
This requires even more skilled workers to perform many different tasks in widely diverse contexts to meet market requirements. The human workforce is integrated with the manufacturing systems, and people too need to be flexible and adaptive \citep{Yew2016}. 
AR technology, as highlighted in this paper, provides remarkable tools 
and features to support the needs of this new workforce, by leveraging on and enhancing the human cognitive abilities towards augmenting their abilities and productivity.

In this paper, we survey studies related to corporate training, including training programs provided by organisations to empower the workforce to meet client expectations and to contribute to both the profitability and the mission of the company. 

AR technologies are scrutinised to identify the main benefits as measured due to their application. Furthermore, we try and expose the disadvantages noted by authors. 
Moreover, to verify the current trend regarding the number of articles published, and check on business acceptance of AR for training purposes, we classify surveyed articles regarding AR configurations, quality of results, tracking methods used and the display hardware employed in experiments.


\section{Methodology}

The first step when conducting a systematic literature review is to define the research objectives and circumscribe the problem being addressed \citep{tenorio2016does}. This systematic literature review assesses AR as used to promote work performance. This paper aims to answer the following research questions:

\begin{enumerate}
 \item Is AR attracting more interest to be used as a corporate training recently?
 \item  What are the main challenges to adopting AR in OJT? 
 \item  What are the main benefits achieved by AR in OJT? 
 \item  Is AR a potencial tool to be used for OJT? 
 \end{enumerate}
 
Before starting the systematic study we need to define a search strategy for primary publications, documenting empirical studies germane to our questions~\citep{keele2007guidelines}. 

We adopted a two-stage strategy: first we defined the keywords and the semantics of research and then we selected which digital libraries, journals, and conferences to search for studies considering the following factors:

\begin{itemize}
 \item Availability of articles in bibliographic databases;
 \item Search availability from keywords to define a search string 
 \item Relevance of bibliographic database \citep{tenorio2016does,keele2007guidelines}.
\end{itemize}

To define the search strings to be explored, the authors discussed and agreed on 
(1) specifying a keyword string and (2) analyse, observe and evaluate the search results from the same databases for both its relevance and its representativeness.

Based on the possible results and to structure the systematic review, the selected string  should feature a sufficiently large corpus of articles and 
their content should make it possible to answer the proposed research questions. 
After discussing the possible results within the authors and theirs limitations, the selected result string includes all articles and journals that feature the words ((“AUGMENTED REALITY”) OR (“MIXED REALITY”)) AND (TRAINING) as keywords or in their titles. We chose five different databases, considering the previously discussed criteria: ACM Digital Library, IEEE Xplore, Elsevier (Science Direct), Elsevier Scopus, and SpringerLink.

Aiming to improve the results, we defined different selection criteria (inclusion, exclusion, and quality) based on the research question, String search and bibliographical databases. Our objective was to identify primary papers that would provide direct evidence about the research questions, also to reduce the likelihood of bias~\citep{keele2007guidelines}. Specifically, we used the guidelines for performing systematic literature reviews in software engineering \citep{keele2007guidelines} and PRISMA (Preferred Reporting Items for Systematic Reviews and Meta-Analyses) \citep{prisma1,prisma2} as a methodological foundation for the selection, evaluation, and exclusion phases.

As a means to facilitate this understanding, we summarise our findings in Table~\ref{table-criteria}. 
We read the paper titles to reduce article count. After ward, we evaluated the abstracts for each initially retained article. 
Last, we read the introduction and conclusions of each paper. After applying these filters, we read the remaining papers integrally. To assess the quality of the finally chosen articles, we used ten criteria which we present in Table~\ref{table-summary}. We describe them succinctly below: 

\begin{description}

    \item [ \textbf{RAT}] Is there a rationale 
    and discussion on the assumptions upon which the study was 
    based? Does 
    the study feature a specific goal to be achieved in the AR scenario? \citep{tenorio2016does}
    \item [ \textbf{LL}]  
    Are the study conclusions grounded on empirical research or does it feature a “lessons learned” report based on expert opinion? \citep{tenorio2016does}
    \item [ \textbf{OBJ}]  Is there a clear statement of the goals of the research? This is important to ascertain to what extent they were met \citep{tenorio2016does}
    \item [ \textbf{MAT}]  Are the materials and research methods clearly described? A clear and detailed description would make it easier to better assess and reproduce the study results. 
    \item [ \textbf{CTXT}]  Is there an adequate description of the context (such as industry, laboratory setting, and products used) in which the research was carried out? Again, a clear description of context would support third party verification and reproduction of the study.
    \item [ \textbf{EMP}]  Was the study empirically evaluated and its results quantified? An empirical evaluation would strengthen the conclusions of a given study.
    \item [ \textbf{DISC}]  Is there a discussion about the results of the study and its impacts? A clear and reasoned argument on the validity and generalisability of the conclusions significantly adds to the contribution of the report.
    \item [ \textbf{LIM}]  Are the limitations of this study explicitly discussed? Such a discussion provides greater insights on the follow-up research and greatly adds to the contribution of the study as reported \citep{tenorio2016does}. 
    \item [ \textbf{OUTC}]  Does the research also add value to the industrial community? Does the paper conclusions apply to, or identify 
    process improvements? Can these be extrapolated beyond the original scope?
    \item [ \textbf{AR}]  Are the display and tracking techniques clearly described? A clear description allows for better contextualisation and assessment of outcomes. 
\end{description}

Each paper received one point for each 
criterion duly addressed. During this reading, we collected data to provide the information needed for this 
classification. Furthermore, we only retained articles that scored a five minimum point grade for this SLR.


\begin{table}[]
\fontsize{8}{8} \selectfont

\caption{Summary of inclusion and exclusion criteria.}
\resizebox{\textwidth}{!}{%
\begin{tabular}{ll}
\hline
Inclusion                                                                                                                    & Exclusion                             \\ \hline
\begin{tabular}[c]{@{}l@{}}Features a training scope based on a tool that uses AR \\ as its main characteristic\end{tabular} & Academic training                     \\
Published from January 2014 to August 2020                                                                                   & Presents less than 5 pages            \\
English is the primary language                                                                                              & Duplicated articles                   \\
Full articles                                                                                                                & Augmented virtuality training focused \\ \hline
\end{tabular}%
}
\label{table-criteria}

\end{table}

We searched each of the databases separately seeking articles from January 2014 to August 2020 that would match our seed terms. We thus collected 1952 candidate publications, including duplicates. It is important to note that we restricted our search to manuscripts written in English. Articles written in other languages did not contribute to those initial 1952 texts.

We removed 85 duplicate articles, reducing the total number to 1867. After that, all the scanning through titles and keywords have been done, and then articles identified as irrelevant ones were excluded. 
This step was performed by one of us and double-checked by another to avoid biased information or any misinterpretation trimming the total candidate texts from 1867 to 310.

At least one author read the abstract and classified each of the the remaining essays. 
This classification and any doubts that arose were discussed within the group. We excluded 
pieces not compliant with the scope of work, reducing the SLR to 180 entries. 

\begin{figure}[hbt]
\centering

{\includegraphics[scale=0.7]{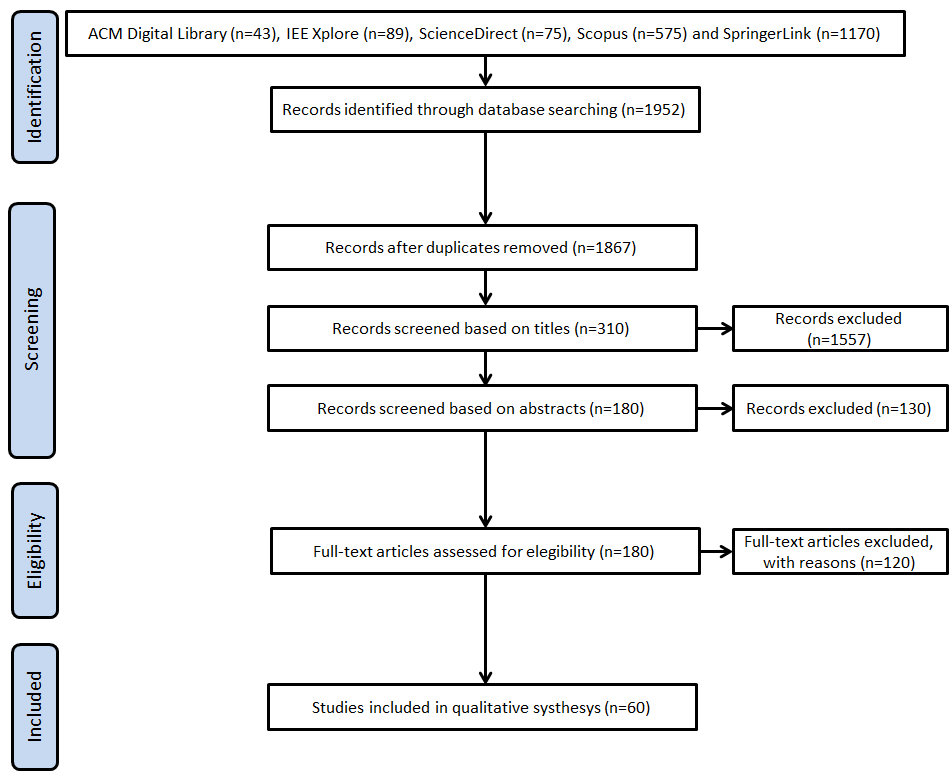}}
\caption{PRISMA flow diagram: study selection process.} \label{figure1}
\end{figure}

From this point on, we retrieved each complete paper from the corpus and one of us read both its introduction and conclusions to eliminate 
non-relevant studies. 
The reader shared the classifications and we discussed possible doubts 
reducing the corpus to 60 papers. Figure~\ref{figure1} depicts the selection process.

Finally, the remaining 60 papers were read in their entirety by 
two authors. We retrieved the data assessing the quality of the 60 selected
publications to verify and illustrate their relevance according to the selected parameters. 
Any doubts or disagreements were shared within the group to guarantee consistency in the final information. The parameters identified and a summary of studies included in the SLR are presented in Table~\ref{table-summary}. As the papers were read, all the relevant information was registered, in order to fulfil this systematic review goal and to structure the data to be analysed.

\begin{center}

\begin{table*}[!ht]

\addtolength{\tabcolsep}{-3pt}
\setlength{\tabcolsep}{4pt}
\fontsize{7}{8} \selectfont

\caption{Summary of studies included in the SLR and the parameters used.}
\centering
{\begin{tabular}{lllllllllllllll}
\toprule
Reference  &
RAT
&
LL
&
OBJ
&

MAT
&
CTXT
&
EMP
&
DISC
&
LIM
&
OUTC
&
AR
&
Score
\\ \midrule
\citep{wright2017design}          & + & + & + & + & + & + & + & + &   & + & 9\\
\citep{martino2017enedis}         & + & + & + & + & + & + & + &   & + & + & 9 \\
\citep{rogado2017evaluation}          & + & + & + & + & + & + & + & + & + & + & 10\\
\citep{torres2018experiences}  & + & + & + & + & + & + & + &   &   & + & 8 \\
\citep{pena2018furthering}       & + & + &   & + & + & + & + & + &   & + & 8\\
\citep{abhari2014training}          & + & + & + & + & + &   & + & + &   & + & 8 \\
\citep{Li2018}              & + & + & + & + & + & + & + &   &   &  & 7 \\
\citep{hou2017framework}             & + & + & + & + & + & + & + & + & + & + & 10 \\
\citep{ramirez2015application}         & + & + & + & + & + & + & + & + & + & + & 10 \\
\citep{Ullo2019}           & + & + & + & + & + &   &   & + & + &  & 7 \\
\citep{mendoza2015augmented}        & + & + & + & + & + & + & + &   & + &  & 8  \\
\citep{mourtzis2018augmented}        & + & + & + & + & + & + & + & + & + & + & 10\\
\citep{hovrejvsi2015augmented}               & + & + & + & + & + & + & + & + & + & + & 10\\
\citep{jetter2018augmented}          & + & + & + & + & + & + & + & + & + &   & 9\\
\citep{syberfeldt2016dynamic}      & + & + & + & + & + & + & + & + & + & + & 10 \\
\citep{borsci2015empirical}          & + &   & + & + & + & + & + & + &   &  & 7 \\
\citep{quandt2018general}         & + & + & + & + & + &   & + & + & + & + & 9\\
\citep{aebersold2018interactive}       & + & + & + & + & + & + & + & + &   &  & 8 \\
\citep{perdikakis2015introducing}      & + & + & + & + & + & + & + & + &   & + & 9\\
\citep{segovia2015machining}         & + & + & + &   & + & + & + &   & + & & 7  \\
\citep{sorkoa2019potentials} & + & + & + &   & + &   & + &   &   &  & 5  \\
\citep{longo2017smart}          & + & + & + & + & + & + & + & + & + & & 9  \\
\citep{tatic2017application}        & + & + & + & + & + &   &   &   &   & + & 6\\
\citep{barsom2016systematic}          & + &   & + &   & + &   & + & + &   &  & 5 \\
\citep{sebillo2016training}         & + &   & + & + & + &   & + & + &   & + & 7\\
\citep{Westerfield2015}     & + & + & + & + & + & + & + & + & + & + & 10\\
\citep{Uva2018}             & + & + & + & + & + & + & + & + & + & + & 10\\
\citep{doshi2017use}          & + & + & + & + & + & + & + &   & + & + & 9\\
\citep{piedimonte2018applicability}    & + & + & + &   & + &   & + & + &   &  & 6 \\
\citep{Lee2019}                   & + & + & + & + & + & + & + & + & + & + & 10\\
\citep{Wang2018}            & + &   & + &   & + &   & + &   &   & + & 5\\
\citep{Stefan2018}          & + & + & + & + & + & + & + & + &   &  & 8 \\
\citep{Bacca2018}          & + & + & + & + & + & + & + & + &   & + & 9 \\
\citep{Kobayashi2018}       & + & + & + &   & + &   & + & + & + & + & 8\\
\citep{limbu2018ar}           & + & + & + & + & + & + & + & + & + & + & 10\\
\citep{Stone2017}          & + & + & + & + & + &   & + & + &   & + & 8\\
\citep{rochlen2017first}         & + & + & + & + & + & + & + & + &   & + & 9\\
\citep{okazaki2017override}   & + & + & + & + & + & + & + & + & + & + & 10\\
\citep{mitsuhara2017using}       & + & + & + & + & + & + & + & + &   & + & 9\\
\citep{Tamaazousti2015}     & + & + & + &   & + &   & + & + & + & + & 8\\
\citep{Herron2016}                 & + &   & + &   & + &   & + & + &   &  & 5 \\
\citep{Wang2016}           & + & + & + &   & + & + & + & + & + &  & 8 \\
\citep{Bifulco2014}         & + & + & + & + & + & + & + & + &   & + & 9 \\
\citep{leitritz2014critical}        & + & + & + & + & + & + & + & + &   & + & 9\\ 

\citep{kwoncomparative}        & + & + & + & + & + & + & + &  &   & + & 8\\

\citep{yang2019augmented}        & + & + & + & + & + & + & + & + & +  & +  & 10\\

 \citep{buttner2020augmented}        & + & + & + & + &  & + & + & + & + & + & 9\\

\citep{koo2019combining}        & + & + & + & + & + & + & + & + &   & + & 9\\

\citep{ferrati2019developing}        & + & + & + & + & + & + & + & + & +  & + & 10\\

\citep{van2020developing}        & + & + & + & + & + & + & + & + & +  & + & 10\\

\citep{catal2019evaluation}        & + & + & + & + &  & + & + & + & +  & + & 9\\

\citep{balian2019feasibility}        & + & + & + & + & + & + & + & + &   & + & 9\\

\citep{wang2020information}        & + & + & + & + & + & + & + & + & +  & + & 10\\

\citep{interactive19}        & + & + & + & + & + & + & + &  &   & + & 8 \\

\citep{pilati2020learning}        & + & + & + & + &  & + & + & + & +  & + & 9\\

\citep{aziz2020mixed}        & + & + & + & + & + & + & + &  & +  & + & 9\\

\citep{eder2020application}        & + & + & + & + & + & + & + & + & +  & + & 10\\

\citep{koutitas2020performance}        & + & + & + & + & + & + & + & + &   & + & 9\\

\citep{romero2019training}        & + & + & + & + &  & + & + & + & +  & + & 9\\

\citep{gabajova2019virtual}        & + & + & + & + &  & + & + &  & +  & + & 8\\

\bottomrule
\end{tabular}
}
\label{table-summary}

\end{table*}

\end{center}


\section{Results and Discussion}

\begin{figure}[hbt]
\centering
\subfloat[Final publication distribution by country.]
{
\resizebox*{9cm}{!}{\includegraphics{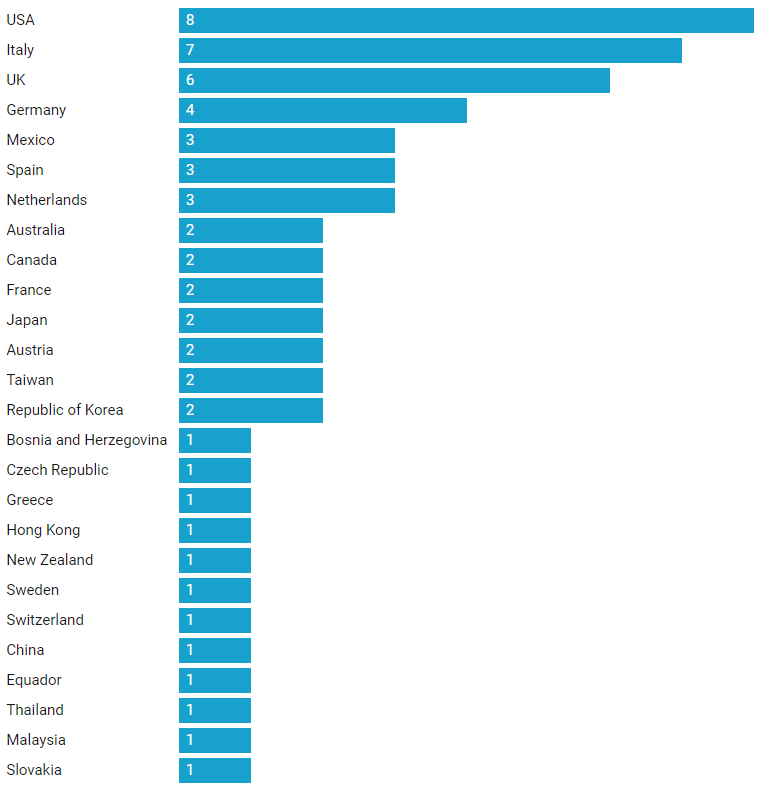}}}\hspace{5pt}
\subfloat[Final publication distribution by continent.]{
\resizebox*{9cm}{!}{\includegraphics{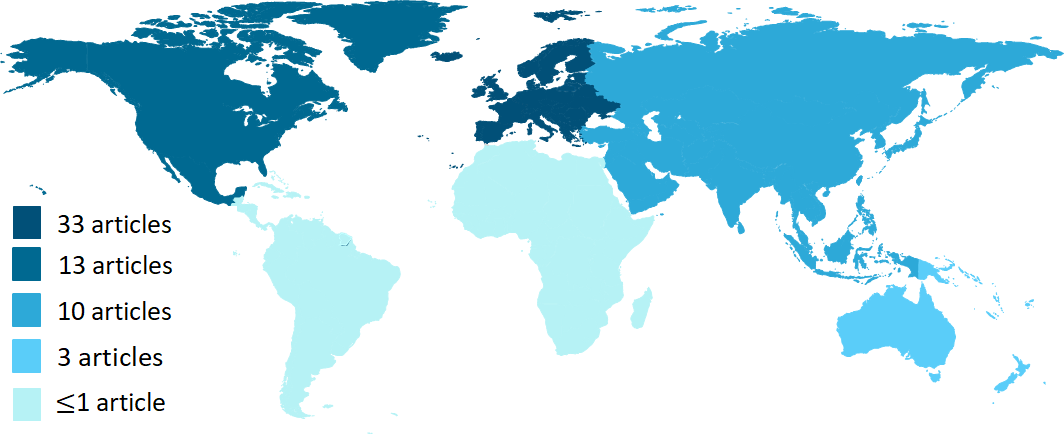}}}
\caption{Final publication distribution by continent and by country.} \label{figure2}
\end{figure}

After reading and analysing each of the remaining 60 papers, we extracted and translated the data into information in a process guided by answer based on the research questionnaires described previously. This will be detailed in the following subsections. Figure~\ref{figure2}(a) shows for each paper its country of origin. Notably, only twelve papers come from outside Europe or North America. USA and Italy are responsible for 15 manuscripts. European researchers alone produced 34 studies as can be seen on Figure~\ref{figure2}(b). This analysis is relevant to identify the locations of research centres focusing on AR for corporate training. 
Thus, it becomes easier to highlight regions where AR in corporate training is most relevant. Therefore, we can conclude that developed countries contribute a larger share than developing countries, suggesting that investments to perform AR research are more prone to be made in locations with more resources and more educated workers.

\begin{figure}[hbt]
\centering

{\includegraphics[scale=0.7]{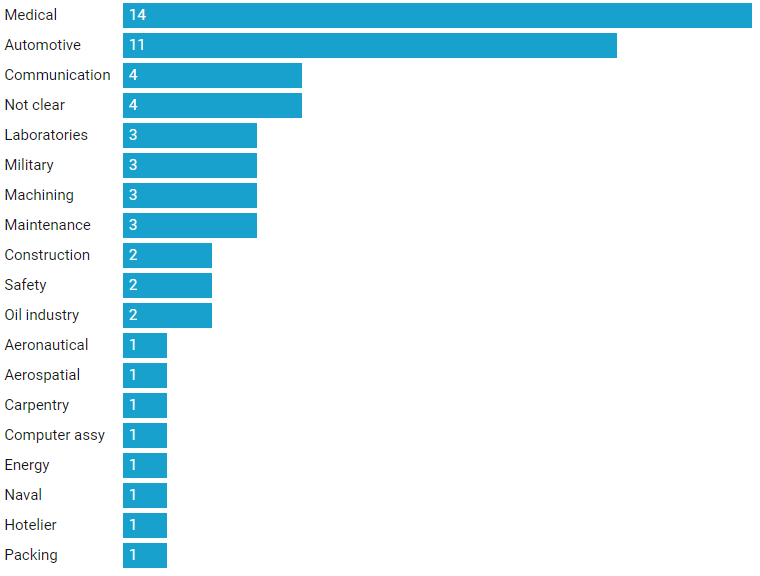}}
\caption{Final area distribution.} \label{figure3}
\end{figure}

Figure~\ref{figure3} identifies industries where AR is applied. Indeed, the automotive industry has the potential to become the principal application for AR training in manufacturing environments, featuring 11 papers. Another issue of note is that AR applications supporting blended learning for medical training have garnered both public and scientific interest as they are the subject of 13 publications. 
The remaining studies are distributed more evenly among diverse economic sectors.

\begin{figure}[hbt]
\centering
\subfloat[Final display classification.]
{%
\resizebox*{8.5cm}{!}{\includegraphics{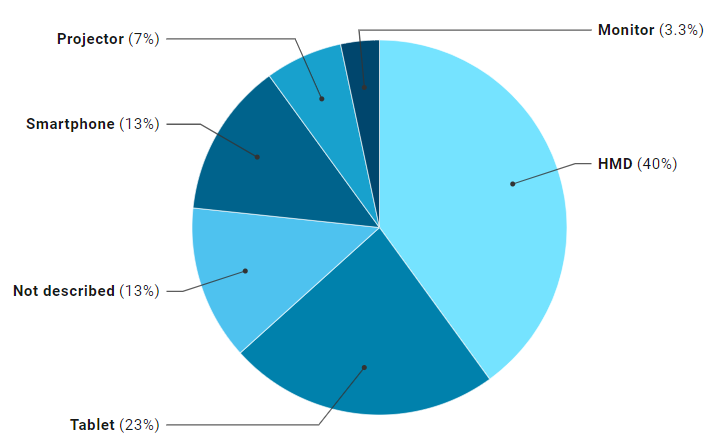}}}\hspace{5pt}
\subfloat[Final tracking method classification.]{
\resizebox*{9cm}{!}{\includegraphics{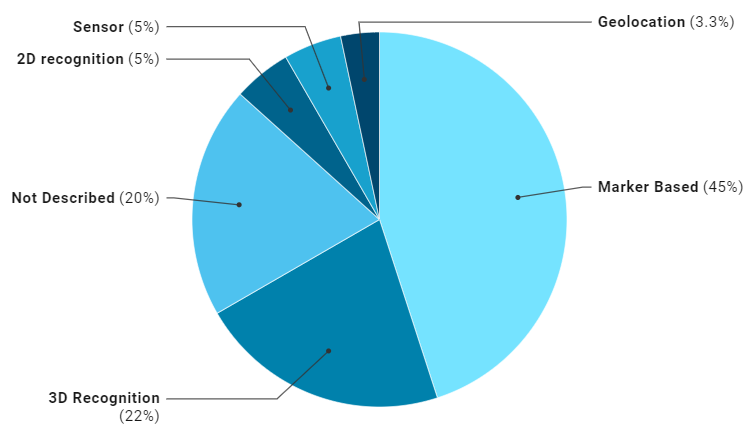}}}
\caption{Final display and tracking method classification.} \label{fig:displays}
\end{figure}

Figure~\ref{fig:displays} shows 
display configurations used in research as well as the tracking methods applied.
As we can see, HMDs are most commonly used, as they free both hands to execute required tasks during the training activities. A tablet is the second most used display both due its flexibility and its low cost. Finally, projectors and monitors make for a smaller share of display technologies. Although the use of HMD leads the applications, the sum of non-immersive displays is superior than immersive ones  Further analysis of display technologies can be found later in this section. 

Marker-based tracking methods are seemingly the most popular. Indeed, marker-based tracking is a well-established mechanism in AR \citep{schmalstieg2011augmented} and its popularity may be due to the reduced cost \citep{pinto2008bratrack} besides that cameras can detect 
fiducial markers in real time with no difficulty ~\citep{pfeiffer2014eyesee3d}.

Although 12 publications do not clearly describe the tracking method adopted, the data are still valid to 
list 3D recognition in second place, with an increasing trend for the near future. 15 papers feature a more computationally demanding technology, markerless tracking, that combines 2D and 3D recognition 
Sensor-based approaches are less popular and include electromagnetic and inertial tracking.

The following subsections 
discuss the four Research Questions (RQ) in detail, discussing the relevant findings in the surveyed literature.

\subsection{RQ1 -  Is AR attracting more interest to be used as a corporate training recently?}

\begin{figure}[hbt]
\centering

{\includegraphics[scale=0.7]{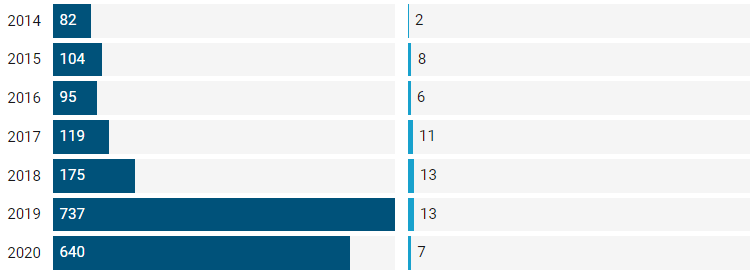}}
\caption{Final publication year distribution.} \label{fig:year} 
\end{figure}

As it is observed in Figure~\ref{fig:year}, there is a steady increase in publications on corporate training over the past few years. Indeed, from the initial year considered in this SLR until the entire last year, their number has grown significantly. 
Also worth noting, the number of publications featuring AR and training in either their title or keywords has steadily increased over the years, illustrating a growing interest of the scientific community.

Putting into perspective, the percentage of studies regarding corporate training has been stable from 2.43\% to 1,75\% in recent year , following the increasing trend in the AR articles published. While in 2014 the selection rate was 2/82, in 2019 this number reached 13/737. 
Thus AR is increasing as the main tool in training, both causing and reflecting the growing number of publications studying its effectiveness. 
Thus we can answer positively Research Question 1:  Is AR attracting more interest to be used as a corporate training recently. 
Moreover, training is a growing area of application for AR, taking into account the share of training related papers among AR scientific publications.

\subsection{RQ2 - What are the main challenges to AR in OJT?}

Only 16 of the selected articles do not address difficulties in developing training applications. On the other hand, the remaining 60, identify at least one challenge. 
In general, the following issues have been reported on using AR as a training tool: adapting non-tech savvy users to AR technology \citep{wright2017design}, training engagement \citep{rogado2017evaluation}, field of vision \citep{Wang2018}, visual occlusion limitations \citep{Herron2016}, ergonomics \citep{Stone2017}, environment interference \citep{Bifulco2014}, dependence on internet connection \citep{Ullo2019}, dependence on batteries, fear of changes \citep{pena2018furthering}, handling perspective and depth \citep{abhari2014training}, management engagement \citep{Li2018}, high cost of 
customization \citep{longo2017smart}, synchronizing reality  and virtuality \citep{hou2017framework}, choosing the training scope \citep{ramirez2015application}, resource costs when updating AR training content \citep{Ullo2019}, acquiring trainers for interpersonal interaction \citep{mourtzis2018augmented}, camera focus \citep{hovrejvsi2015augmented}, software issues \citep{Ullo2019} and ergonomic issues of wearables when used continuously \citep{jetter2018augmented}.

Training people non-familiar with the technology was a challenge identified in eight of the studies including~\cite{perdikakis2015introducing}. Since AR training systems commonly use software and require minimum digital expertise, people experiencing difficulties with IT are either prone to reject AR, be too ashamed to ask for help, or do not engage in the experience~\citep{wright2017design}. The interest in and value perception of AR technologies are likely to be higher for both students and technology-savvy people and may also vary according to gender~\citep{Habig2019}.

Another issue regarding engagement is percentage of studies that highlighted either low enthusiasm or a low perception of benefits by trainers. Indeed, 8.3\% of the studies discussed this finding. Notably, several studies indicate that either the managers and the corporate directors actively support and endorse the new training technology or the trainers/operators will tend to reject it. 

Another aspect covered by the analysis is the hardware required. Displays, if not chosen carefully, can pose significant obstacles to training. 
In effect, 12 studies mentioned troubles with the displays selected including~\cite{Herron2016}. The problems either related to the wrong choice of display (ex: instead of using a handheld display, a wearable would fit the function better) or troubles pertaining to the hardware proper, or reported discomfort with wearable devices during continued use. 

Other issues mentioned in the literature include display price point. Since many expensive displays are exclusively used for specialised niche training applications, they might not be cost-effective.

Ten studies reported issues affecting AR markers. 
Both complexity of marker technology and interference from the environment are the leading causes of problems reported in~\cite{Westerfield2015}. 
Placing markers in corners or in open environments can cause misreadings~\citep{Bifulco2014}. Also, the cost and the environmental 
constraints cameras to establish reliable tracking were also discussed. On the other hand, software can also be a source of complications. 
Eleven studies mentioned the software presenting at least one problem, as discussed by~\cite{hovrejvsi2015augmented}. In this case, the main issue 
was due to inflexible configurations or applications designed for on rails execution and not being able to handle unusual conditions or inputs in a graceful way. 
Other problems reported off-the-shelf software lacking customisation for AR, such as video calling applications. 
Customisation was pointed out as a strong feature in 18 articles e.g.~\citep{torres2018experiences} given the added flexibility. However,
eight papers identify room for improvement as seen in the~\cite{longo2017smart}. 
Customising software can be difficult mainly due to the deep knowledge required to edit the training software, or the preparation need to change 
training sequences or dialogues. Even though when inside a training session it may be possible to insert additional information or change 
feedback triggered from markers relatively easy and cheap, changing the
business logic of the training software or physically moving tracking devices for a different training module, can be much harder and expensive.

Internet access is flagged as an issue in two studies~\citep{pena2018furthering,Ullo2019}, mainly due to unstable connections, or insufficient bandwidth as AR requires video streaming when 
deployed remotely, or when it requires external processing.

The 3D perception and perspective are mentioned as challenges by 14 studies
namely~\cite{Kobayashi2018}. Properly matching virtual content with real settings to meet training requirements is essential to avoiding perception troubles. 
Simulator (3D) sickness~\citep{Lee2019} needs to be avoided, as the distortion between real objects and virtual content minimises breaks in the flow of work and increases the trainee's attention to the instructional content. 

Table~\ref{table-challenges} synthesises the main points extracted during
our survey regarding challenges to the adoption of AR. We also describe them succinctly below: 

\begin{description}
    \item [ \textbf{LELP}] Low enthusiasm or a low perception of benefits by trainers; in many contexts this may mean scepticism towards unproven technologies or a steep learning curve characteristic of some early systems; 
    \item [ \textbf{EIFC}] Engagement issues or fear of changes; Indeed in many corporate environments, introducing a new technology can be met with resistance by people who perceive their job security being threatened; 
    \item [ \textbf{3DP}] 3D perception regarding the perspective and deepness; These maybe related to display quality, registrationb problems or both;
    \item [ \textbf{HWI}] Hardware issues; Some of the early hardware could be quirky and prone to failure; 
    \item [ \textbf{DMR}] Difficulties regarding the marker readings; These might either be caused by poor illumination or unfortunate choice of surfaces;
    \item [ \textbf{DBC}] Difficulties with broadband connection and data transfer; Perceived lag can cause simulator sickness, poor registration or even missing key animations; network latency can exacerbate these issues;
    \item [ \textbf{CI}] Customisation issues; As AR training environments can be complex, budgetary issues can lead to under-featured poorly adapted software; 
    \item [ \textbf{SOFT}] Software issues; Bugs or poorly documented features can lead to unanticipated behavior; 
    \item [ \textbf{TPNF}] Training people non-familiar with the technology; Some more complex systems may have a steep learning curve, which prevents non-specialised people to take full advantage of AR for training of job-specific tasks;
    \item [ \textbf{NDNC}] Not described or not clear.

\end{description}

\begin{center}

\begin{table*}[!ht]
\caption{Summary of challenges identified in the SLR}

\addtolength{\tabcolsep}{-3pt}
\fontsize{8}{8} \selectfont

\centering
{\begin{tabular}{lllllllllll}
\toprule
Reference
& 
LELP
&
EIFC
&
3DP
&
HWI
&
DMR
&
DBC 
&
CI
&
SOFT
&
TPNF
&
NDNC

\\ \midrule
\citep{wright2017design}           & + &   &   &   &   &   &   &   &   &   \\
\citep{martino2017enedis}          &   &   &   &   &   &   &   &   &   & + \\
 \citep{rogado2017evaluation}           &   & + &   &   &   &   &   &   &   &   \\
 \citep{torres2018experiences}   &   &   &   &   &   &   &   &   &   & + \\
 \citep{pena2018furthering}        &   & + & + & + & + & + &   &   &   &   \\
 \citep{abhari2014training}           &   &   & + &   &   &   &   &   &   &   \\
 \citep{Li2018}               &   & + &   &   &   &   &   &   &   &   \\
 \citep{hou2017framework}              &   &   &   &   &   &   & + & + &   &   \\
 \citep{ramirez2015application}          &   &   &   &   &   &   & + & + &   &   \\
 \citep{Ullo2019}             &   &   &   &   &   &   & + &   &   &   \\
 \citep{mendoza2015augmented}          &   &   &   &   &   &   &   &   &   & + \\
 \citep{mourtzis2018augmented}         &   & + &   &   &   &   &   &   &   &   \\
 \citep{hovrejvsi2015augmented}                 &   &   &   &   & + &   &   & + &   &   \\
 \citep{jetter2018augmented}           & + & + & + & + &   &   & + &   &   &   \\
 \citep{syberfeldt2016dynamic}       &   &   & + & + &   &   & + &   &   &   \\
 \citep{borsci2015empirical}           &   &   &   &   &   &   &   &   & + &   \\
 \citep{quandt2018general}           &   &   &   &   & + &   &   &   &   &   \\
 \citep{aebersold2018interactive}        &   &   &   &   &   & + &   &   &   &   \\
 \citep{perdikakis2015introducing}       & + &   &   &   & + &   &   & + &   &   \\
 \citep{segovia2015machining}          &   &   &   &   &   &   &   &   &   & + \\
 \citep{sorkoa2019potentials}  &   &   &   &   &   &   &   &   &   & + \\
 \citep{longo2017smart}            &   &   &   &   &   &   & + &   &   &   \\
 \citep{tatic2017application}         &   &   &   & + &   &   &   &   &   &   \\
 \citep{barsom2016systematic}           &   &   &   &   & + &   & + & + &   &   \\
 \citep{sebillo2016training}          & + &   &   &   &   &   &   &   &   &   \\
 \citep{Westerfield2015}      &   &   & + &   & + &   &   &   &   &   \\
 \citep{Uva2018}              &   &   &   &   &   &   & + &   &   &   \\
 \citep{doshi2017use}            &   &   &   &   &   &   &   &   &   & + \\
 \citep{piedimonte2018applicability}     &   &   &   &   &   & + &   & + &   &   \\
 \citep{Lee2019}                     &   &   & + &   &   &   &   &   &   &   \\
 \citep{Wang2018}             &   &   & + & + &   &   &   &   &   &   \\
 \citep{Stefan2018}           &   &   &   &   &   &   &   &   &   & + \\
 \citep{Bacca2018}            &   &   &   &   &   &   &   &   &   & + \\
 \citep{Kobayashi2018}        &   &   & + &   &   &   &   &   &   &   \\
 \citep{limbu2018ar}            &   & + & + &   & + &   &   &   &   &   \\
 \citep{Stone2017}            &   & + & + &   &   &   &   &   &   &   \\
 \citep{rochlen2017first}          &   &   &   & + &   &   &   & + &   &   \\
 \citep{okazaki2017override}    &   & + &   &   &   &   &   &   &   &   \\
 \citep{mitsuhara2017using}        &   &   &   &   &   &   &   & + &   &   \\
 \citep{Tamaazousti2015}      &   &   & + &   &   &   &   &   &   &   \\
 \citep{Herron2016}                  &   &   & + & + &   &   &   &   &   &   \\
 \citep{Wang2016}             &   &   &   &   &   &   &   &   &   & + \\
 \citep{Bifulco2014}          &   &   & + &   & + &   &   & + &   &   \\
 \citep{leitritz2014critical}         &   &   &   &   &   &   &   &   &   & + \\

 
 \citep{kwoncomparative}        &  &  & + &  &  &  &  &  &  &  \\

\citep{yang2019augmented}        &  &  & + &  &  &  &  & + &  &  \\

 \citep{buttner2020augmented}        & + &  &  &  &  &  &  &  & + &  \\

\citep{koo2019combining}        &  &  &  &  &  &  &  &  &  & + \\

\citep{ferrati2019developing}       &  &  &  & + & + &  &  &  &  &  \\

\citep{van2020developing}        & + & + &  & + &  &  &  & + & + &  \\

\citep{catal2019evaluation}      &  &  &  &  &  &  &  &  &  & + \\

\citep{balian2019feasibility}      &  &  &  &  &  &  &  &  & + &  \\

\citep{wang2020information}      &  &  &  &  &  &  &  &  &  & + \\

\citep{interactive19}        &  &  &  & + &  &  &  &  &  &  \\

\citep{pilati2020learning}       &  &  &  &  &  &  &  &  &  & + \\

\citep{aziz2020mixed}      &  &  &  &  &  &  &  &  &  & + \\

\citep{eder2020application}       &  &  &  & + &  &  &  &  &  &  \\

\citep{koutitas2020performance}    &  &  &  & + &  &  &  &  & + &  \\

\citep{romero2019training}       &  &  &  &  & + &  &  &  & + &  \\

\citep{gabajova2019virtual}       &  &  &  & + &  &  &  &  &  &  \\
 

 \bottomrule
\end{tabular}
}
\label{table-challenges}

\end{table*}

\end{center}

\subsection{RQ3 - What are the main benefits achieved by AR in OJT?}

From examining the 60 articles we conclude that AR, although with different descriptions, provides the following advantages: reduced training costs  \citep{Ullo2019}, facilitated customization  \citep{Uva2018}, raised effectiveness~\citep{Doshi2016}, low-risk when exercising critical safety issues~\citep{rogado2017evaluation}, attractive to specific groups~\citep{wright2017design}, flexible information display~\citep{pena2018furthering}, improved worker confidence~\citep{torres2018experiences}, fast access to information~\citep{sebillo2016training}, support to decision-making~\citep{Kobayashi2018}, improved skill transfer process~\citep{pena2018furthering}, real-time interaction~\citep{sebillo2016training}, empowering operators~\citep{syberfeldt2016dynamic}, displaying immersive environments~\citep{Li2018}, familiarisation with the work routine~\citep{abhari2014training}, allowing non-specialised staff to perform specific tasks~\citep{Ullo2019}, manpower savings~\citep{hou2017framework}, decreased training time~\citep{Wang2016}, decreased perceived distances~\citep{hou2017framework}, decreased error rates~\citep{leitritz2014critical}, reduced cognitive workload~\citep{okazaki2017override}, easy to store and transport~\citep{perdikakis2015introducing}, decreased set-up time~\citep{quandt2018general}, welcome by users~\citep{mitsuhara2017using}, increased motivation~\citep{Bacca2018}, friendly remote assistance  and better long term retention of information~\citep{Kobayashi2018} among others. In this subsection, we will discuss how the previously mentioned benefits are documented throughout the articles and analyse how these affect the perceived suitability of AR to training in the working environment. Although 15 articles presented at least a quantified benefit, 59 mentioned qualitatively perceived or immeasurable gains. 


Tallying the information from~\citep{ramirez2015application,mourtzis2018augmented,hovrejvsi2015augmented,longo2017smart,Westerfield2015,Uva2018,Lee2019,ferrati2019developing,koutitas2020performance} time savings with respect to traditional approaches average 28.48\% (considering only the final number provided in the respective paper). 
An additional 27 articles mentioned unquantified gains related to time spent as compared to traditional methods. The main issue regarding time measurements is that the training suite can advance stages as the operator progresses, as all modules are geared to individual learning and the information presented is focused, in order to narrow attention to the training goals. Another characteristic observed is that training modules can use display resources to provide unsolicited information at any moment. Effectively, there is no need to wait for human-generated events to happen, as the simulation can evolve in real time. Another key benefit highlighted by studies is the perceived lower error rate during
task execution when using AR. In effect, \cite{mourtzis2018augmented,Uva2018,Doshi2016,leitritz2014critical, ferrati2019developing,koutitas2020performance,balian2019feasibility} measured the operator errors to be, on average, about 47.85\% lesser than conventional settings during the task execution. 20 articles mentioned a decrease in related mistakes in comparison with traditional methods. 
AR can provide specific, relevant and customised information about every step of the training, which can explain why AR-based training 
tends to yield less mistakes. Another corroborating characteristic is that AR presents information graphically and thus makes it easier to understand than text-based delivery.

Cost-related issues are also relevant. Besides the inherent money savings due to learning time reduction and decreased error rates, the costs regarding learning content production can be addressed. One study~ \cite{ramirez2015application} measured these to be 41\% lesser as compared to traditional methods. Besides that, 25 studies mentioned, without measuring, that total cost of training was lower, namely~\cite{Stone2017}.

An advantage that is immeasurable at this point but worth mentioning is 
time savings by senior operators cited by 4 essays ~\citep{Ullo2019,ferrati2019developing,wang2020information,interactive19}  in the production environment who would otherwise be engaged to help and provide information to new employees.

The flexibility afforded by AR was both welcome and noticed in 19 articles, including ~\cite{torres2018experiences}, which stated that AR makes 
content customisation easier. Once the main body and training methodology are defined, it becomes simpler to change environments~\citep{sebillo2016training}, change the information given, or even update these in real-time depending on perceived requirements.

The possibility to simulate a given environment is quoted by 25 studies
e.g.~\citep{barsom2016systematic}. Also, improvements in safety are quoted by 7 papers e.g.~\citep{rogado2017evaluation,torres2018experiences,Li2018,barsom2016systematic,wang2020information}. These two characteristics are strongly related. As it becomes easier to simulate an environment, specific conditions can be simulated too~\citep{aebersold2018interactive}. Hence it is possible to train the operator using comparatively inexpensive equipment without incurring health or safety risks in case of misguidances~\citep{Stone2017}. It is possible to simulate events and manage stressful situations without incurring extra costs beyond content production. Dangerous environments or unsafe activities can also be simulated without major safety issues to the operator, as discussed in ~\cite{Stefan2018}, it is not necessary to expose novice operators to radiation, but training is required on how to manage emergency situations in hazardous environments e.g. nuclear power plants.

The cognitive advantages of AR are discussed in 29 articles. 
Indeed, many operators welcome and experience positive attitudes towards AR-based learning as it requires less cognitive resources to achieve 
results comparable to traditional methods ~\citep{okazaki2017override}. A key advantage is that learning scenarios can be repeated as many times as 
needed without embarrassment.  
Since the delivery is strongly visual the scene can be placed exactly at the time and place relevant to training, avoiding possible confusion. The same situation can also be played in different ways, as a poka-yoke\footnote{Poka-yoke is a Japanese term that means "mistake-proofing" or "inadvertent error prevention". A poka-yoke is any mechanism in any process that helps an equipment operator avoid (yokeru) mistakes (poka).} to overcome any misunderstanding of the message the training is supposed to impart~\citep{jetter2018augmented}. Table~\ref{table-benefits} synthesises the main points extracted during our literature survey regarding the main benefits outlined above.  We also describe them briefly below: 

\begin{description}
    \item [ \textbf{RTC}] Reduced training cost;
    \item [ \textbf{ECE}] Easy customisation and editing;
    \item [ \textbf{LER}] Lower error rate during task execution;
    \item [ \textbf{LTT}] Lower training time;
    \item [ \textbf{LCL}] Lower cognitive load;
    \item [ \textbf{ARFT}] AR provided environment facilities for training;
    \item [ \textbf{SAFTI}] Safety improvements;
    \item [ \textbf{BNDNC}] Benefits not described or not clear.
     
\end{description}

\begin{table*}[!ht]
\caption{Summary of benefits of AR-based learning and training as identified in our literature review}

\fontsize{8}{8} \selectfont


\centering
{\begin{tabular}{lllllllll}
\toprule
Reference
&
RTC 
&
ECE 
&
LER 
&
LTT 
&
LCL 
&
ARFT 
&
SAFTI
&
BNDNC

\\ \midrule
 \citep{wright2017design}          & + & + &   &   &   &   &   &   \\
 \citep{martino2017enedis}         & + & + & + & + & + &   &   &   \\
 \citep{rogado2017evaluation}          &   &   &   &   &   & + & + &   \\
 \citep{torres2018experiences}  &   & + &   & + & + &   & + &   \\
 \citep{pena2018furthering}       & + & + &   &   & + &   &   &   \\
 \citep{abhari2014training}          &   &   &   & + & + & + &   &   \\
 \citep{Li2018}              & + &   &   &   & + & + & + &   \\
 \citep{hou2017framework}             & + &   &   & + &   & + &   &   \\
 \citep{ramirez2015application}         & + &   &   & + &   &   &   &   \\
 \citep{Ullo2019}            & + &   &   & + &   &   &   &   \\
 \citep{mendoza2015augmented}         & + &   & + & + &   &   &   &   \\
 \citep{mourtzis2018augmented}        &   &   & + & + &   &   &   &   \\
 \citep{hovrejvsi2015augmented}                & + & + &   & + &   &   &   &   \\
 \citep{jetter2018augmented}          &   &   & + & + & + & + &   &   \\
 \citep{syberfeldt2016dynamic}      & + & + & + & + & + &   &   &   \\
 \citep{borsci2015empirical}          &   &   &   &   &   &   &   & + \\
 \citep{quandt2018general}          & + & + & + & + &   & + &   &   \\
 \citep{aebersold2018interactive}       &   &   &   & + & + & + &   &   \\
 \citep{perdikakis2015introducing}      & + & + &   &   & + & + &   &   \\
 \citep{segovia2015machining}         & + &   &   &   &   &   &   &   \\
 \citep{sorkoa2019potentials} & + &   & + & + &   & + &   &   \\
 \citep{longo2017smart}           &   &   &   & + & + & + &   &   \\
 \citep{tatic2017application}        &   &   & + &   &   &   &   &   \\
 \citep{barsom2016systematic}          &   & + &   &   & + & + & + &   \\
 \citep{sebillo2016training}         &   & + &   & + & + &   &   &   \\
 \citep{Westerfield2015}     & + &   &   & + &   &   &   &   \\
 \citep{Uva2018}             & + & + & + & + & + &   &   &   \\
 \citep{doshi2017use}           & + &   & + &   &   &   &   &   \\
 \citep{piedimonte2018applicability}    & + & + & + &   & + &   &   &   \\
 \citep{Lee2019}                    & + &   &   & + & + &   &   &   \\
 \citep{Wang2018}            & + &   & + & + &   &   &   &   \\
 \citep{Stefan2018}          &   &   & + & + &   & + &   &   \\
 \citep{Bacca2018}           &   &   & + &   & + &   &   &   \\
 \citep{Kobayashi2018}       &   & + &   & + & + & + &   &   \\
 \citep{limbu2018ar}           &   & + &   &   & + &   &   &   \\
 \citep{Stone2017}           & + &   &   &   &   & + &   &   \\
 \citep{rochlen2017first}         &   &   &   &   & + & + &   &   \\
 \citep{okazaki2017override}   &   &   &   &   & + &   &   &   \\
 \citep{mitsuhara2017using}       &   &   &   &   & + & + &   &   \\
 \citep{Tamaazousti2015}     &   &   &   &   & + & + &   &   \\
 \citep{Herron2016}                 & + &   & + & + &   &   &   &   \\
 \citep{Wang2016}            & + &   &   & + & + & + &   &   \\
 \citep{Bifulco2014}         &   &   &   & + & + &   &   &   \\
 \citep{leitritz2014critical}        &   &   & + &   &   &   &   &   \\

 
 \citep{kwoncomparative}        &  &  &  &  & + &  &  &  \\

\citep{yang2019augmented}        &  &  & + & + &   & + & + &  \\

 \citep{buttner2020augmented}        &  &  & + &   &  & + &  &  \\

\citep{koo2019combining}        &  & + &  &  & + &   &  &  \\

\citep{ferrati2019developing}       & + &  & + & + & +  &  &  &  \\

\citep{van2020developing}        &  & + &  &  &  &   & + &  \\

\citep{catal2019evaluation}      &  &  &  &  & +  & + & + &  \\

\citep{balian2019feasibility}      & + &  & + &  & + & +   &  &  \\

\citep{wang2020information}      &  &  &  &  & + &    &  &  \\

\citep{interactive19}        & + &  &  &  & +  & + &   &  \\

\citep{pilati2020learning}       &  &  &  & + &  &   &  &  \\

\citep{aziz2020mixed}      &  &  &  &  & + &  &   &  \\

\citep{eder2020application}       &  & + &  &  & +  &  &  &  \\

\citep{koutitas2020performance}    & + & + & + & + &  & + &  &  \\

\citep{romero2019training}       &  & + &  &  & + &   &  &  \\

\citep{gabajova2019virtual}        &  & + &  & + & + &   &  &  \\
 
 \bottomrule
\end{tabular}
}
\label{table-benefits}

\end{table*}


\subsection{RQ4 - Is AR a potential tool to be used for OJT?}

Based on the answers to the three previous research questions, 
it is possible to conclude AR has potential to become a useful tool for OJT in many settings. Its features and benefits seemingly mesh well with OJT requirements, and we can see that once development environments support enough flexibility~\citep{Tamaazousti2015}, training suites are geared towards the workplace experience ~\citep{barsom2016systematic}, the learning experience becomes job-oriented, learning module contents can be updated with relative ease and training packages can be used without specialised supervision~\citep{Ullo2019}. The just-in-time based e-learning can be supported by AR technology since the training modules just require the operator to be available and the training station prepared~\citep{mendoza2015augmented}.

The advantages highlighted in the RQ3 fit well with most objectives of OJT. 
Indeed, 59 studies report strong benefits as exemplified by~\cite{quandt2018general}, and only 46 described difficulties, notably in~\cite{syberfeldt2016dynamic}. Nonetheless, as far as the result we analysed, 15 of these studies could quantify productivity gains e.g.~\citep{leitritz2014critical}. On the other hand, none of the reported disadvantages could be precisely measured during the published experiments. 
While current trends show AR as being applied to individual training tasks, 
comprehensive OJT can be the next application to enjoy widespread AR deployment. 

\subsection{Limitations of the study}
Even though the authors intended to cast as wide a net as possible, the articles reviewed in this research are limited to the journals indexed in the databases: ACM Digital Library, IEEE Xplore, Elsevier (Science Direct), Elsevier Scopus, and SpringerLink.  
This research is limited to experimental studies presented in article format, thus possibly missing more anecdotal or corporate proprietary evidence. Although all selected studies were integrally read, some info could be missing or misunderstood during the analysis. As the field evolves and given the rapid pace of AR development, future iterations of this study could shed more light on our findings.
In addition to that, this SLR is limited to academic papers in a review intended to assess the imminent impact of a technology in the industry.


\section{Conclusion}


From the analysis of the benefits and disadvantages presented in this paper, it is clear that AR is a maturing approach to corporate training scenarios as the technology overcomes its growing pains and associated challenges. 
Indeed, at the current development stage, well-thought-out planning stages before training programs and learning content are deployed can maximise both skill improvement and workforce quality. This planning should include the correct evaluation of the target audience. Indeed,  
depending on the digital skill levels of the target audience, 
AR applications can be complicated and feature unnecessary layers or be straight and simple to address less sophisticated or casual audiences.

The display selection can be critical to success and should consider both the training scope, environmental, safety, and ergonomics issues. If the activity requires a hands-free operation, we recommend either an HMD or a projector-based setting. If the training is time-intensive, 
wearable devices can present ergonomic issues. If the budget does not include specific hardware for each worker, or if the training has to be conducted simultaneously by many trainees, it should be desirable to adopt smartphones or tablets as primary displays. 


Tracking methods need attention while considering the environment.
Variations in lighting can interfere with the reading of markers depending on the camera used. The surface on which to place the markers should be studied. If irregular surfaces are likely to cause misreadings, other tracking devices such as a sensor-based or a markerless tracker could be chosen.

The training environment should be correctly prepared to support any flow of action that the trainees shall undergo. Avoiding inflexible trainee-"on rails" experiences should improve the learning experience. 

Software development should take into account the ultimate training objectives and accommodate the stakeholders mentioned above.

If developers are to heed these recommendations, they can increase the training's benefits and minimise the challenges of using AR as an OJT useful tool.





Following current trends, design thinking should contribute better results to developing AR training software. Another issue to take into account is gamification\cite{barata2013} as a device to optimise and to engage trainees. Moreover, just-in-time solutions can facilitate software development regarding both the game engine and training structure to decrease efforts required to develop the software and increase the application range of the training and its access to more developers.


Although the studies identified different benefits and some of them performed measurements, there is still a strong need for a comprehensive survey that combines all measurements and compares them with the traditional OJT training methods. The challenge consists of AR-based training in an industrial OJT application for the mastering of one activity, in which the following critical performance indicators shall be directly compared with traditional training programs: (1) the cost, (2) the training pace, (3) the time required, (4) the learning curve, (5) the cognitive workload, (6) the perception regarding AR and (7) the technology acceptance.


As a means to develop future work relevance, AR will increasingly use smartphones or other wearable devices to accommodate the growing adoption of AR everyday settings and improve people's familiarity with the technology. However, AR can still evolve in fundamental ways to improve both the user and learning experiences. Further Artificial Intelligence developments can make for more natural and sophisticated dialogues and better coaching of trainees to identify possible pitfalls and proactively suggest better ways to accomplish tasks. Further advances in HMDs and wearable devices will surely make AR applications usable for extended periods, a typical worker complaint since the inception of the field. Finally, ever more advanced user interface techniques, coupled with more natural interaction modalities going beyond speech and gestures, should make for ever more natural user experiences, increasing worker engagement, and productivity of AR-based learning environments.

\section*{Acknowledgements}


The last author would like to thank the São Paulo Research Foundation (FAPESP) for support of this research: grant\#2018/20358-0, FAPESP.

This work was partially supported by Portuguese government national funds through FCT, \textit{Fundação para a Ciência e a Tecnologia}, under project UIDB/50021/2020.






\section*{Notes on contributors}



\emph{Bruno R. Martins} is a student at the Professional Master in Technological Innovation, Federal University of Sao Paulo, where he is researching about the augmented reality applications on training. He is graduated in industrial engineering and simultaneously works as an engineer at EMBRAER S.A on the manufacturing system. His main research interests are Virtual Reality, Augmented Reality and On-The-Job Training.


\emph{Joaquim A. Jorge} is a full professor of computer graphics and multimedia at the Department of Computer Science and Engineering of the IST, University of Lisbon, Portugal, and scientific coordinator of the Research Group on Visualisation and Multimodal Interfaces of the Institute for Computer and Systems Engineering - INESC-ID. Since 2007 he is editor in chief of Computers \& Graphics Journal and serves or has served on the board for six other international journals.

\emph{Ezequiel R. Zorzal} is an associate professor of computer science \& engineering at Institute of Science and Technology, Federal University of Sao Paulo. He is an integrated researcher at Professional Master in Technological Innovation. He is simultaneously an associate researcher of the Visualisation and Intelligent Multimodal Interfaces Group at INESC-ID, IST, University of Lisbon, Portugal. His main research interests are virtual reality, augmented reality, educational and medical user interfaces.







\bibliographystyle{apalike}
\bibliography{article}

\end{document}